\documentclass[prl,aps,epsfig,twocolumn,showpacs,floatfix]{revtex4}
\usepackage{graphicx}
\usepackage{color}
\usepackage{epsfig}

\begin{document}
\title{Controlling the Ratchet Effect for Cold Atoms}
\author{Anatole Kenfack$^{(a)}$, Jiangbin Gong$^{(b)}$, and
Arjendu K. Pattanayak$^{(c)}$}

\date{\today}

\affiliation{(a) Max-Planck-Institut f\"ur Physik Komplexer Systeme,
N\"othnitzer Strasse 38, D-01187 Dresden, Germany\\
(b) Department of Physics and Center for Computational Science and Engineering,
National University of Singapore, 117542, Republic of Singapore\\
(c) Department of Physics and Astronomy, Carleton College, Northfield,
Minnesota 55057, USA}
\begin{abstract}
Low-order quantum resonances manifested by directed currents have
been realized with cold atoms. Here we show that by increasing the
strength of an experimentally achievable delta-kicking ratchet
potential, quantum resonances of a very high order may naturally
emerge and can induce larger ratchet currents than low-order
resonances, with the underlying classical limit being fully chaotic.
The results offer a means of controlling quantum transport of cold atoms.
\end{abstract}

\maketitle

The ratchet effect, i.e., the possibility to derive directed
transport without bias in periodic systems with broken symmetries,
was originally proposed by Feynman. This effect, prohibited in
systems at equilibrium by the second law of
thermodynamics~\cite{feynmann1966}, has recently gained renewed
interest~\cite{haenggi2002,reimann2002} as a model for the physics
of molecular motors~\cite{juelicher1997}. Directed transport is
possible when particles are driven out of equilibrium and relevant
spatio-temporal symmetries are broken~\cite{flach2000}. This has
motivated the construction of nanoscale devices in which artificial
ratchets may serve as new electrons pumps, molecular switches and
particle selectors, among other applications~\cite{juelicher1997,
astumian1997,lehmann2002}. Other studies have shown that when the
noise is absent, its role can be replaced by deterministic chaos
induced by the inertial term~\cite{jung1996}. In such inertial
ratchets, the issue of current reversal was intuitively
addressed~\cite{mateos2000} and later carefully
reformulated~\cite{kenfack2007}. Purely Hamiltonian ratchets, where
noise and friction are eliminated, have received notable attention
as well~\cite{flach2000,dittrich2000}.

Besides these classical ratchets,  quantum Hamiltonian ratchet
effects arising from purely unitary evolution is also possible.
These are very important, for example, for the design of coherent
nanoscale devices~\cite{reimann1997}. Exploring quantum coherence
phenomena in chaotic Hamiltonian ratchets hence becomes necessary.
The quantum delta-kicked rotor (QKR), a paradigm of quantum
chaos~\cite{casati1995}, is a convenient and experimentally
realizable model for such explorations, possessing dynamical
localization~\cite{fishman1982}, quantum accelerator
modes~\cite{schlunk2003}, tunneling~\cite{reimann1997,linke1999}, as
well as quantum
resonances~\cite{shepelyanskii1980,moore1995,daley2002,duffy2004,
sandro2004, dana2006,ryu2006}. The ``quantum ratchet accelerator",
where the coherent ratchet current accelerates linearly, was first
studied with a modified QKR and later in the kicked
Harper model~\cite{gong200406}.

Since the pioneering experiment of cold-atom
ratchets~\cite{mennerat-robilliard1999}, new designs looking for
ratchet effects in nonlinear Hamiltonian systems
~\cite{monteiro2002} have emerged. Motivated by the first
experimental
 realizations of sawtooth-like asymmetric
potentials \cite{ritt2006,salger2007} as well as quantum resonance
ratchets~\cite{sadgrove2007}, in this Letter we revisit the quantum
flashing ratchet model in Ref.~\cite{lundh2005}, with the
perspective of detecting and controlling quantum resonance dynamics
of very high orders.  The ultimate goal is to help design powerful
means for the coherent control of the dynamics of cold atoms with
driven but dissipationless optical lattices. For other cold-atom
control scenarios using also Hamiltonian ratchet effects, see
Refs.~\cite{luis2007,gong2007}.

A quantum resonance occurs when the flashing period is commensurate
with the recoil frequency and is related to the arithmetic nature of
the effective Planck constant $\widetilde{\hbar}$ of kicked systems,
occuring specifically if
\begin{equation}
\label{Eq:resonance}
\widetilde{\hbar}=4\pi r/s,
\end{equation}
with $r,s$ being mutually prime integers. Cases with small $s$ and
large $s$ values can be called low-order quantum resonance (LOQR)
and high-order quantum resonance (HOQR), respectively. We show below
that HOQRs can manifest themselves strongly in the ratchet current
behavior, with their corresponding classical phase space being fully
chaotic. This further enhances the view that quantum control
techniques can be applied to classically chaotic systems~\cite{gongreview}.

The system we consider is described, in dimensionless units, by  the
following Schr\"odinger equation~\cite{lundh2005}
\begin{equation}
i\widetilde{\hbar}\frac{\partial \psi}{\partial
t}=-\frac{\widetilde{\hbar}^2}{2}\frac{\partial^2\psi}{\partial
x^2}+ v(x)\sum_{l=0}^{\infty}\delta(t-l)\psi\label{schroedinger}
\end{equation}
where $x$ is the position, and the potential $v(x)=K[\sin(x)+\alpha
\sin(2x)]$ is assumed to be periodically flashed off and on with
delta kicks. Here $t$ is the time variable and $l$ an integer that
counts the number of kicks. By superimposing a conventional standing
wave potential of $\lambda/2$ spatial periodicity with a
fourth-order lattice potential of $\lambda/4$ periodicity, such a
dissipationless ratchet potential $v(x)$ has been successfully
engineered~\cite{ritt2006,salger2007}. The scheme, satisfying the
Raman-Nath transition processes~\cite{ritt2006,salger2007,chabe2007},
uses three level atoms with two stable ground states and one electronically
excited state (for more experimental details, see Ref.~\cite{ritt2006}).
Our results
below can thus be experimentally verified by taking parameters from
Refs.~\cite{ritt2006,salger2007} for alkali-metal atoms such as Rb
and Cs. Note that $\alpha=V_2/V_1$ and $K=V_1/2$, where $V_1$ and
$V_2$ denote the potential depths of the lattice harmonics
$\lambda/2$ and $\lambda/4$, respectively.
The parameter $\alpha$
controls the skewness of the potential. With $\alpha \in (0,0.5]$
the sawteeth of the potential lean left, stimulating the transport
to the right for classical diffusive motion. The familiar rotor
potential can be recovered for $\alpha=0$. For this temporally-symmetric
system, the ratchet effect is only possible for broken spatial symmetry,
i.e., for $\alpha\neq 0$. We use scaled units here with both the spatial
$L$ and the temporal $T$ periods set to unity. The quantum nature of the
system is in the effective Planck constant
$\widetilde{\hbar}=8\omega_RT$, which varies as one adjusts the
pulsating period $T$. Here $\omega_R=\hbar k_L^2/2m $ is the
recoil frequency of the applied laser field, with $m$ the atom mass
and $k_L$ the photon wave number that makes up a lattice period of
$(2k_L)^{-1}$ for the optical potential.

The quantum map of the above delta-kicked ratchet model is given by
$ \hat{U}=\exp(-i \widetilde{\hbar} \hat{k}^2/2)\exp(-iP
v(\hat{x}))$, where $\hat{x}$ and
$\hat{k}=-i\frac{\partial}{\partial x}$ represent the position and
the wave number operators, respectively. Here we have also defined
$P = K/\widetilde{\hbar}$ for later use. All computational examples
presented below  are for fixed skewness parameter $\alpha=0.3$.

Quantum resonances have been long studied in kicked rotor
systems~\cite{shepelyanskii1980,daley2002, moore1995,duffy2004,
sandro2004}. A direct observation of quantum resonance dynamics is
known to require initial states with long coherence width. This
issue is heightened for HOQRs. Very recently non-condensed
atoms were used to indirectly observe a particular family of
HOQRs~\cite{kanem2007}. However, using Bose-Einstein condensates
loaded in optical lattices~\cite{ryu2006,sadgrove2007}, one can now
easily realize initial quantum states whose coherence spans many
optical lattice sites, resulting in the observations of the main
quantum resonance~\cite{sadgrove2007} as well as quantum resonances
of relatively low orders~\cite{ryu2006}. Given this experimental
progress we assume below, unless stated otherwise, that the initial
wavefunction is homogeneous with zero initial momentum. We then let
$\widetilde{\hbar}$ take specific values and explore new quantum
resonance effects in the ratchet transport.

\begin{figure}[htb]
\epsfig{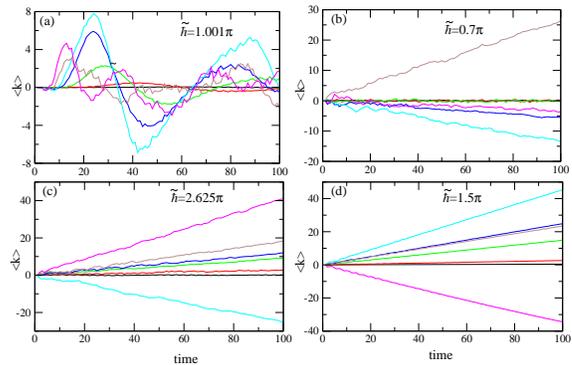}
\caption{\label{figure1}Time dependence of the ratchet current for
different values of $P$ and $\widetilde{\hbar}$. In panels (a), (b)
and (d), $P=$ $0.5$ (black), $1.0$ (red), $2.0$ (green), $3.0$
(blue), $4.0$ (cyan), $5.0$ (brown) and $6.0$ (magenta). (a)
$\widetilde{\hbar}=1.001\pi$ and no directed transport occurs for
any value of $P$. (b) $\widetilde{\hbar}=0.7\pi$ with the tuple
$(r,s)=(7,40)$; transport occur in either direction, at
$P$-dependent acceleration rates.  (d) As in (b) except
$\widetilde{\hbar}=1.5\pi; (r,s) =(3,8)$. In panel (c) $P= 1.0$
(black), $2.0$ (red), $3.0$ (green), $4.0$ (blue), $5.0$ (cyan),
$7.0$ (brown) and $8.0$ (magenta), with
$\widetilde{\hbar}=2.625\pi$; $(r,s)=(21,32)$. (Color online)}
\end{figure}

The ratchet current, denoted $\langle k\rangle$ below, is defined as
the expectation value of $\hat{k}$.
Its time dependence is shown in Fig.~(1) for several values of the
potential strength $P$ and for specific values of
$\widetilde{\hbar}$. One clearly sees in Fig.~\ref{figure1}(a) that
there is no directed transport for $\widetilde{\hbar}=1.001\pi$ with
any value of the potential strength.  By sharp contrast, the current
acceleration may be strongly favored in one or the other direction
as illustrated in Fig.~\ref{figure1}(b), (c), and (d), where
$\widetilde{\hbar}=0.7\pi$ for $(r,s)=(7,40)$,
$\widetilde{\hbar}=2.625\pi$ for $(r,s)=(21,32)$, and
$\widetilde{\hbar}=1.5\pi$ for $(r,s)=(3,8)$, respectively. Note
that cases in Figs.~2(b) and 2(c) represent quantum resonances of
very high orders (much higher than those observed in
Ref.~\cite{ryu2006}). Unlike the ratchet current of a low order
resonance (where current reversal may be well understood by noticing
that the wavefunction amplitude at a point $x$ at $t=l$ is a
coherent sum of the wavefunction amplitudes at other locations at
$t=l-1$~\cite{lundh2005}), here a simple explanation of the HOQR
current dependence on $P$ is not available. Indeed,  an analytical
treatment of the ratchet transport associated with those very high
order resonances seems very difficult, even for a perturbation
theory using very small $\alpha$~\cite{dario2007}.

\begin{figure}[htb]
\epsfig{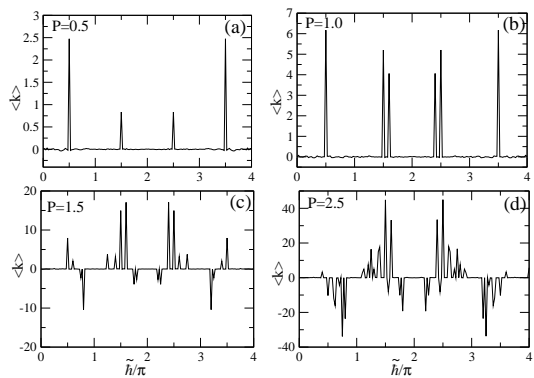}
\caption{\label{figure2}Ratchet current $\langle k \rangle$ as a
function of $\widetilde{\hbar}/\pi$ after 200 kicks, with the
potential parameter $\alpha=0.3$, and the potential strength $P$
indicated in each panel. Main resonances appear in (a) with low $P$.
As $P$ increases in (b), (c), and (d), full chaos is being developed
(see Fig. 3) and significant ratchet currents due to higher-order
quantum resonances emerge.}
\end{figure}

Figure \ref{figure2} displays $\langle k\rangle$ after $200$
temporal periods as a function of $\widetilde{\hbar}$. In Fig. 2(a)
$P=0.5$ and the results correspond to the findings of
Ref.~\cite{lundh2005}, showing a net drift at those quantum
resonances where $\widetilde{\hbar}/\pi$ is given by half-integers.
As we increase $P$ the system exhibits dramatic changes. For
example, when $P=1.0$ in Fig.~\ref{figure2}(b) two twin peaks
emerge. One also sees, as shown in Fig.~\ref{figure2}(c) and (d),
that larger values of $P$ lead not only to a proliferation of peaks,
but also to current reversals. As shown in Table I, many of these
peaks are found to be associated with very high-order quantum
resonances covering a wide range of $(r,s)$. Remarkably, these
high-order resonances may yield larger ratchet current acceleration
than the main resonances, and in either case the current direction
depends on $P$. Note, however, that HOQRs do not always transport
better than LOQR (compare, for example, the HOQR at $\hbar/\pi=0.6$
to the LOQR at $\hbar/\pi=0.5,1.5,3.5$). Figure 2 also shows that
relatively small changes in $\widetilde{\hbar}$ can dramatically
change the ratchet current, thus experimentally offering a means of
isolating different HOQRs. This also suggests that particles with
slightly different masses, hence slightly different
$\widetilde{\hbar}$ and different $P$ due to an isotope effect, may
display qualitatively different kinds of transport. Neglecting at
the moment the non-ideal situation in experiments, note that the
resonance peaks shown in Fig.~(2) are better resolved with
increasing kicks since the absolute amplitude of the current peaks
is proportional to the number of kicks.

\vspace{0.4cm}
\begin{tabular}{c|| c | c|c|c|c|c}
\hline \multicolumn{7}{c} {Table I: Some high-order resonances
in Fig.~\ref{figure2}(d)}\\
\hline\hline
$\widetilde{\hbar}/\pi$&$0.6$&$0.7$&$0.75$&$1.125$&$1.55$&$3.3$\\
\hline
$(r,s)$&(3,20)&(7,40)&(1,16)&(9,32)&(31,80)&(33,40)\\
\hline
\end{tabular}\\
\vspace{0.4cm}

We now comment on the possibility of experimentally observing these
HOQR peaks. Using constraints associated with state-of-the-art
experiments, we have carried out extensive
computations~\cite{kgpinprep}, verifying that the above-observed
HOQR ratchet current can be clearly observed in experiments, at
least for time scales of $20-30$ kicks. In particular, the ratchet
currents are sufficiently stable when (i) considerable dephasing is
present, (ii) a finite pulse-width instead of delta-kicks is
considered, and (iii) a realistic quasi-momentum spread in the
initial state is considered. For example, for the case of
$\widetilde{\hbar}=2.625\pi$, $P=5.0$ shown in Fig.~1(c), we find
that the HOQR ratchet current deviates considerably from the ideal
case only after about 20 kicks, for a realistic quasi-momentum
spread as estimated in Ref.~\cite{ryu2006}. We have also checked
that if a superposition state of momentum (such as in
Ref.~\cite{sadgrove2007}) is used as the initial state, then a HOQR
ratchet current can be also effectively demonstrated without using a
bichromatic optical lattice.

\begin{figure}[htb]
\epsfig{file=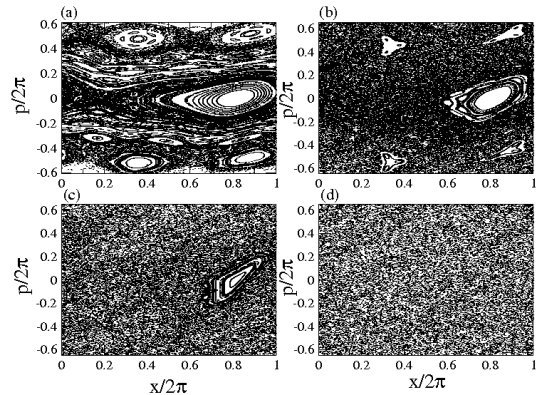,width=7.cm} \caption{\label{figure3}
Classical phase space structures for the kicked ratchet map for
$\alpha=0.3$ showing regular islands embedded in the chaotic sea for
(a) $K=0.25\pi$, (b) $K=0.55\pi$ and (c) $K=0.70\pi$.  In panel (d)
$K=0.8\pi$ and full chaos is reached.}
\end{figure}

Let us now briefly describe the classical dynamics of the model
system. With the classical stochasticity parameter or kick strength
$K=\widetilde{\hbar}P$, the associated classical map is given by $
p_{l+1}=p_l-K [\cos(x_l)+2\alpha\cos(2x_l)];$ $ x_{l+1}=x_l+
p_{l+1}$, where $p_l$ is the momentum variable conjugated to the
coordinate $x_l$. Figure \ref{figure3} displays the classical phase
space for $\alpha=0.3$ and for varying $K$. As $K$ increases, the
islands initially dominating the phase-space shrink and decrease in
number, [see Fig.~\ref{figure3} (a), (b), (c)], until a threshold
value $K_{\text{thr}}$, when essentially full chaos is reached. In
Fig.~\ref{figure3}(d), $K=0.8\pi>K_{\text{thr}}\approx 0.75\pi$, the
entire phase space is seen to be chaotic.

\begin{figure}[htb]
\epsfig{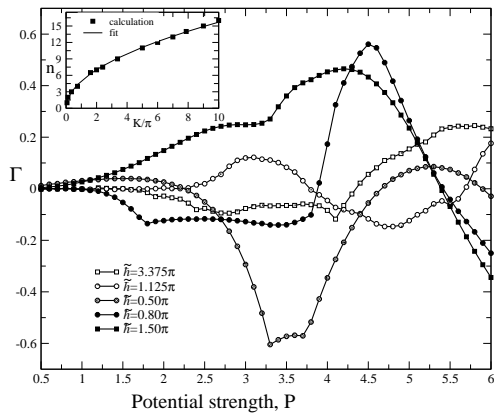} \caption{\label{figure4}
Acceleration rate $\Gamma$ as a complicated function of the potential
strength $P$ for different $\widetilde{\hbar}$. The inset shows the
number of Floquet bands below the potential barrier as a function of
$K$, fitted empirically as $n\propto \sqrt{K}$.}
\end{figure}

Remarkably, we observe that clear HOQR peaks of the ratchet current
emerge only when the classical counterpart is fully chaotic. This
observation has been checked by varying $\alpha$ in the range $0$ to
$10$, with different $K_{\text{thr}}$. Such a connection between a
purely quantum phenomenon and a purely classical phenomenon is
worthy of some remarks, though a profound explanation may not exist.
First, quantum resonances lead to continuous energy
bands~\cite{shepelyanskii1980}. A potential of height $P=K/\hbar$
supports only a certain number of energy bands (denoted $n$) that
are below the potential barrier. As the well deepens, $n$ increases
with $K$, specifically $n \propto \sqrt{K}$ as shown in the inset in
Fig.~\ref{figure4}.  For a fixed $\widetilde{\hbar} = K/P$,
larger $K$ corresponds to more classical chaos, and leads to more
bands that can contribute to transport, whence a HOQR is more likely 
to be detected. Alternatively, as we increase $K$ for fixed
$P$, $n$ increases, leading to more resonant values
$\widetilde{\hbar}$ being available. Both ways, the result is more
peaks in the plot of the ratchet current vs $\widetilde{\hbar}$. As
such, chaos and HOQRs, both requiring sufficiently large $K$, go
hand-in-hand, an interesting result also noticed~\cite{kanem2007}
using other signatures. Since classical chaos arises through the
growth of nonlinear resonances~\cite{reichl}, its connection with
HOQRs might exist at a deeper level, though further work along this
line is beyond the scope of this paper.  Finally, note that for our
Hamiltonian ratchet model, directed transport may occur for $n<2$,
albeit being weaker than in cases with larger $n$. This is in
contrast to dissipative systems~\cite{grifoni2002}, where the
ratchet effect exists only if there are at least two bands below the
barrier.

Finally, to motivate further theoretical work, we show in
Fig.\ref{figure4} the average current acceleration rate
$\Gamma\equiv \langle k\rangle/l$ within $l=100$ kicks, as a
function of $P$, for $\widetilde{\hbar}$ corresponding to a few
HOQRs as well as LOQRs. The size and direction of $\Gamma$ are both
seen to be tunable with $P$, with their $P$-dependence varying
markedly with $\widetilde{\hbar}$.

In summary, we have shown the important role of quantum resonances
of very high orders in a QKR-based quantum ratchet. The associated
transport can be manipulated by use of these high-order quantum
resonances. We also observe, and partially explain, that the
transport associated with HOQRs become important only if the associated
classical phase space is fully chaotic.  The results are of great
experimental interest because (i)~they offer a ratchet acceleration
mechanism previously not noticed and (ii)~suggest a new means of detecting
the intriguing quantum high-order resonances in QKR systems. This study
should help design new means of controlling the dynamics of cold atoms in
pulsed optical lattices.

Fruitful discussions with Kamal P.~Singh and J.-M.~Rost are
acknowledged. A.~K. acknowledges supports from the Alexander von
Humboldt foundation as well as the Max-Planck Gesellschaft through
the Reimar L\"ust fund (2005). J.~G. was supported by the start-up
fund (WBS grant No. R-144-050-193-101 and No. R-144-050-193-133),
and the NUS ``YIA" fund (WBS grant No.: R-144-000-195-123), both from
the National University of Singapore.


\begin{thebibliography}{99}

\bibitem{feynmann1966}
R.~P.~Feynman et al., The Feynman Lectures on
Physics (Addison-Wesley, Reading, MA, 1966), Vol.1, Chap.46.

\bibitem{haenggi2002}
R.~D.~Astumian and P.~H\"anggi, Physics Today {\bf 55} (11), 33
(2002).

\bibitem{reimann2002}
P.~Reimann, Phys.~Rep.~{\bf 361}, 57 (2002).

\bibitem{juelicher1997}
F. J\"ulicher et al., Rev.~Mod.~Phys.~{\bf 69}, 1269 (1997).

\bibitem{flach2000}
S.~Flach et al., \prl{\bf 84}, 2358 (1997); O. Yevtushenko et al.,
Europhys.~Lett.~A{\bf 54}, 141 (2001); R. Gommers et al,
\prl{\bf 96}, 140604 (2006).

\bibitem{astumian1997}
R.~D.~Astumian, Science {\bf 276}, 917 (1997).

\bibitem{lehmann2002}
J.~Lehmann et al., \prl {\bf 88}, 228305 (2002).

\bibitem{jung1996}
P.~Jung et al., \prl {\bf 76}, 3436 (1996).

\bibitem{mateos2000}
J.~L.~Mateos, \prl{\bf 84}, 258 (2000).

\bibitem{kenfack2007}
 A.~Kenfack, S.~M.~Sweetnam, and A.~K.~Pattanayak, \pre{\bf 75} 056215
 (2007).

\bibitem{dittrich2000}
T.~Dittrich et al., Ann.~Phys. (Leipzig) {\bf 9}, 755 (2000); H.
~Schanz et al., \prl {\bf 87}, 070601 (2001); S.~Denisov et al.,
\pre{\bf 66}, 046203 (2002); T.~Cheon et al., J.~Phys.~Soc.~Jpn.
{\bf 72}, 1087 (2003).

\bibitem{reimann1997}
P.~Reimann, M.~Grifoni, and P.~H\"anggi, \prl {\bf 79}, 10 (1997);
P.~Reimann, and P.~H\"anggi, Chaos {\bf 8}, 629 (1998).

\bibitem{casati1995}
{\it Quantum Chaos: Between Order and Disorder}, ed. by G. Casati
and B. Chirikov (Cambridge Univ. Press, Cambridge, England, 1995).

\bibitem{fishman1982}
S.~Fishman et al., \prl {\bf 49}, 509 (1982).

\bibitem{schlunk2003}
S.~Schlunk et al., \prl{\bf 90}, 124102 (2003).

\bibitem{linke1999}
H.~Linke et al., Science {\bf 286}, 2314 (1999).

\bibitem{shepelyanskii1980}
F.~M.~Izrailev and D.~L.~Shepelyanskii, Theor. Math. Phys.{\bf 43},
553 (1980).

\bibitem{daley2002}
A.~J.~Daley and A.~S.~Parkins, \pre {\bf 66}, 056210 (2002).

\bibitem{moore1995}
F.L. Moore et al., \prl {\bf 75}, 4598 (1995).

\bibitem{duffy2004}
G. Duffy et al., \pre {\bf 70}, 056206 (2004).

\bibitem{sandro2004}
S. Wimberger et al., \prl {\bf 92}, 084102 (2004).

\bibitem{dana2006}
I.~Dana and D.~L.~Doforeev, \pre{\bf 73}, 026206 (2006).

\bibitem{ryu2006} C.~Ryu  et al., \prl{\bf 96}, 160403 (2006).

\bibitem{gong200406}
J. Gong and P. Brumer, \pre {\bf 70}, 016202 (2004);
\prl {\bf 97}, 240602 (2006).

\bibitem{mennerat-robilliard1999}
C.~Mennerat-Robilliard et al., \prl {\bf 82}, 851 (1999).

\bibitem{monteiro2002}
T.~S.~Monteiro et al., \prl {\bf 89}, 194102 (2002).

\bibitem{ritt2006} G. Ritt et al., \pra {\bf 74}, 063622 (2006)
\bibitem{salger2007}T. Salger et al., \prl{\bf 99}, 190405 (2007).
\bibitem{sadgrove2007}
M.~Sadgrove et al., \prl {\bf 99}, 043002 (2007); I.~Dana et al,
arxiv: physics/0706.0871.
\bibitem{lundh2005}
E.~Lundh and M. Wallin, \prl {\bf 94}, 110603 (2005).
\bibitem{luis2007} S. Denisov et al., \pra{\bf 75}, 063424 (2007).
\bibitem{gong2007} J. Gong, D. Poletti, and P. Hanggi, \pra{\bf 75},
033602 (2007).
\bibitem{gongreview}J. Gong and P. Brumer, Ann. Rev. Phys. Chem. {\bf
56}, 1 (2005).
\bibitem{chabe2007} J. Chabe et al., arxiV:0709.4320
\bibitem{kanem2007} J.~F.~Kanem et al., \prl {\bf 98}, 083004 (2007).

\bibitem{dario2007}D. Poletti, G.C. Carlo, and B. Li, \pre{\bf 75},
011102 (2007).
\bibitem{kgpinprep}A. Kenfack, J.B. Gong, A.K. Pattanayak, in
preparation.
\bibitem{reichl} {\em The transition to chaos: Quantum manifestations},
L.E.~Reichl, Springer-Verlag, NY, 2004.
\bibitem{grifoni2002}
M.~Grifoni et al., \prl {\bf 89}, 146801 (2002);  J.~B.~Majer et
al., \prl {\bf 90}, 056802 (2003).
\end{thebibliography}
\end{document}